# Urban Morphometrics:
# Towards a Science of Urban Evolution

Jacob Dibble
Urban Design Studies Unit, Department of Architecture, University of Strathclyde, UK
Alexios Prelorendjos
Urban Design Studies Unit, Department of Architecture, University of Strathclyde, UK
Ombretta Romice
Urban Design Studies Unit, Department of Architecture, University of Strathclyde, UK
Mattia Zanella
Department of Mathematics and Computer Science, University of Ferrara,
Emanuele Strano
Geographic Information Systems Laboratory, Ecole Polytechnique Fédérale de Lausanne, CH
Mark Pagel
School of Biological Sciences, University of Reading, UK
Sergio Porta
Urban Design Studies Unit, Department of Architecture, University of Strathclyde, UK

# 1. Introduction
## *1.1. Cities as organisms: beyond the analogical approach*

In a speech delivered before an audience of sustainability scholars in 2004, Christopher Alexander addressed the relevance of an *evolutionary* interpretation of the process of construction, interpreted as a cultural manifestation of the "unfolding" structure of change that is typical of living organisms and nature in general (Alexander, 2004); Alexander argued that the homology between biology and construction must be firmly established at the level of the structure of the generative process, that of *morphogenesis*, rather than of the aesthetics of the final product. This implicit focus on time is an essential feature of everything that is built, but has always been a hostile concept by modern architects and planners, though much more familiar to urban geographers or anthropologists. Among urbanists, urban morphologists are certainly those who peculiarly have placed change – and therefore time – at the heart of their work on the form of cities since the very foundations of the modern discipline (M. R. G. Conzen, 1960; Muratori, 1960); in so doing, they have focused on the component elements of the urban "fabric" rather than the shape of the city as a whole, a focus that we inherit in our study, for example in the choice of the Operational Taxonomic Unit as discussed in the next section. However, after more than a half century since those pioneers' time, a review of the literature in Urban Morphology reveals that the field still lacks a quantitative and universally applicable method for the analysis of urban form. In fact, notwithstanding the remarkable amount of effort spent by the founders themselves, their direct descendants (Cataldi, Maffei, & Vaccaro, 2002; Whitehand, 2001) and international Urban Morphologists across Spain, France, USA, Australia and China (M. P. Conzen, 2001; Darin, 1998; Gu & Zhang, 2014; Ibarz, 1998; Siksna, 2006) amongst others, there is still a lack of a unifying, systematic and quantifiable method of assessing urban form and drawing conclusions from a rigorous analysis of the data. An extensive review of the entirety of the Journal of Urban Morphology and other relevant works in the field reveals that, of the published articles considered, only 23% adopt a primarily systematic approach and only 21% base their conclusions on quantifiable aspects of urban form, and, most importantly, less than 20% do both (Fig.1). We argue that a systematic study of urban form across time deserves more attention, first to understand cities and eventually to act upon them, although this last point is by no means the focus of this paper.

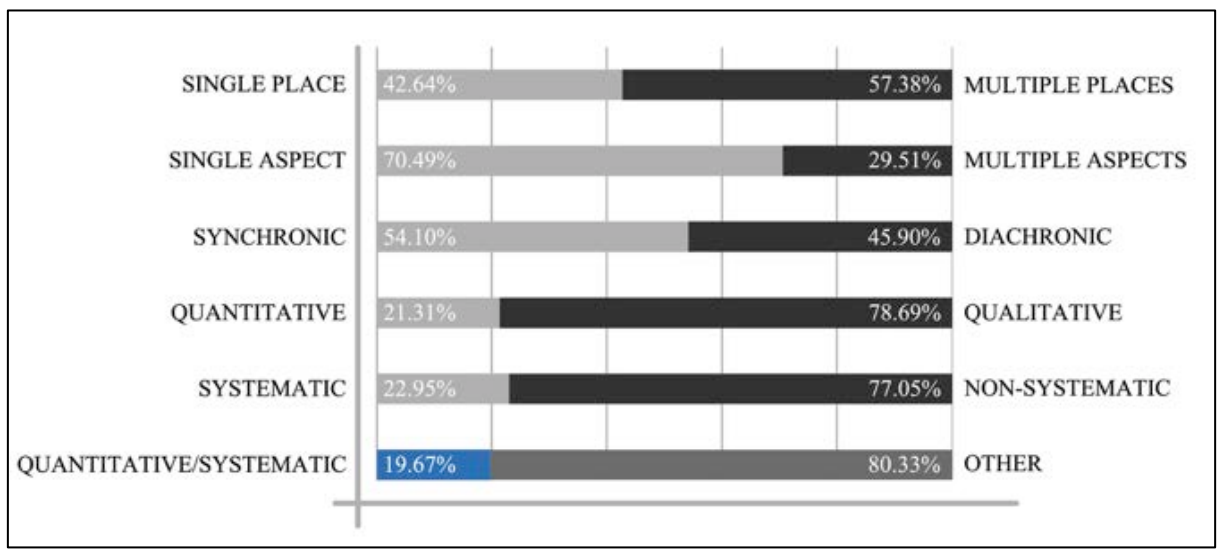

**Figure 1. Trends in urban morphology case study research**
Current case studies research from the last 18 years of the Journal of Urban Morphology as well as other scientific works in the field have been scrutinized according to 5 dichotomies: a) single or multiple places, in that one or multiple locations are considered for analysis; b) single or multiple aspect, in that one or multiple objects of study are utilized for conclusions; c) synchronic or diachronic, if they focus on a single time period or multiple time periods; d) quantitative or qualitative, if the conclusions are based on numerical data or descriptive data; e) systematic or non-systematic, on whether their overall approach reflects a consistent and replicable method. The *systematic* and *quantitative* qualities of the research are most relevant in establishing a science of urban morphometrics. However, of all the scientific works overviewed in the field, less than 20% can claim to exhibit, at any level, both such qualities at the same time.

Our aim in this paper is to test a methodological framework for the systematic investigation of the evolution of urban form. We seek in our work explicit analogies to the evolution of life, moving beyond the metaphorical approaches to cities as mechanisms, organisms or (eco) systems (Marshall, 2008; Steadman, 2008). We view cities as evolved cultural products (Dawkins, 2006, c.1976; Pagel, 2012c; Richerson & Boyd, 2008) whose shapes and forms represent the outcomes of tens of thousands of years of cultural selection for structures that serve basic human needs. To the extent that these needs are universal to our species, we expect similarities in form and function across space and time. On the other hand, a hallmark of human evolution is cumulative cultural adaptation (Pagel, 2012a) characterized by the successive accumulation of technologies and social complexity. One upshot of this is that most humans now live in environments that they could not build on their own, and have little understanding of their workings. To the extent that human needs have themselves changed with these cumulative social and technological changes, we might see new forms and functions emerging over time. However, without further exploring the epistemological and terminological basis of this interpretation of urban evolution, we engage first and primarily in a foundational study of urban morphometrics.

According to Roth and Mercer (Roth & Mercer, 2000) morphometrics in biology is *"the quantitative characterization, analysis, and comparison of biological form"*, which sits at the intersection of developmental and evolutionary biology, i.e. the study of the evolution of developmental mechanisms that drive the growth of living organisms; as such, it is *"a means of extracting information about biological material and biological processes"* (ibidem, p.801). The specification is important, as the study of form is conducive to that of the relationships between organisms, and ultimately to the processes that generate them. The contribution of morphometrics is *"precision in the ability (a) to recognize forms that are intermediate, (b) to judge degrees of proximity or similarity to other forms, and (c) to extrapolate or predict hypothetical, experimental, or nonexistent extremes"* (ibidem, p.802). The *modus operandi* of morphometrics is therefore *"to quantify the size and shape of organisms with the methods of multivariate statistics"* (Klingenberg, 2002, p. 4), in order to shed light on the evolution of forms and in particular on transformations that bring from one form to another (D'Arcy Thompson, 1942, c.1917), where we presume these transformations tell us something about development (ontogeny).

Current scholars in morphometrics distinguish a major shift, or indeed a "revolution" (Rohlf & Marcus, 1993) in the discipline after the introduction, in the 1980s, of a new coherent set of methods operating in particular at the level of form recognition and quantification: "traditional" methods, based on the *algebraic* measurement of distances characterizing the *size* of organs or entire organisms, are now complemented by *geometric* methods that are based on *graphic* processes of recognition and manipulation of their *shape* through the extrapolation of relevant nodes, or *landmarks* (Adams, Rohlf, & Slice, 2013; Reyment, 2010). This new approach can be considered as a synthesis between the two primary traditions of morphometrics: *multivariate biometrics*, emphasizing a focus on the statistical analysis of form rather than geometry, and *geometric visualization*, which focuses on the visible geometric shapes of organisms rather than the numerical quantification of these shapes (Bookstein, 1993). In this perspective, our approach to urban morphometrics opens in the fashion of *traditional morphometrics*: in fact we characterize the form of the urban fabric utilizing a vector of measures that quantify individual aspects of its constituent elements, and indeed their relationship in space.

*1.2. Urban Morphometrics: a systematic understanding of urban form*

Historically, morphometrics has been crucial in building a solid ground for the emergence of evolutionary biology by introducing a rigorous quantification of the phenotypic traits of living organisms and ultimately the analysis of their overall *similarity*. Measurements of morphological traits were instrumental, as well as the consideration of other characteristics, such as behavioral, physiological or molecular aspects, to developing the science of classification that we know under the somehow interchangeable terms of *Taxonomy* or *Systematics* (Manktelow, 2010).

*"Classification is the basic method which man employs to come to grips with and organise the external world. Plants and animals are in fact classified in basically the same way as non-living objects; on the basis of possession of various characters or relations which they have in common"* (Heywood, 1976, p. 1). The

necessity to classify and organise the external world is the fundamental notion of several disciplines of biological and evolutionary sciences, all of which fall under the umbrella category of Systematics. Systematics is the *"scientific study of the kinds and diversity of organisms and of any and all relationships among them"* (Simpson, 1961, p. 7). The result of a Systematics analysis is the derivation of a *system of classification* that best expresses the various degrees of *similarity* between organisms; such systems can be used for the storage, retrieval and communication of information, for facilitating predictions and ultimately for forming generalisations of unknown organisms and inferring relationships between the units that are classified, or *taxa* (Jeffrey & Heywood, 1977).

The concepts and methodologies developed extensively in Systematics are relevant to the rigorous analysis of urban form. In their work on numerical taxonomy, Sneath and Sokal (1973) proceed by first identifying the *Operational Taxonomic Unit* (OTU). The identification of the OTU is a crucial decision that entails the consideration of multiple factors, such as the *purpose* of the classification, the structural organization of what is to be classified, its most appropriate *rank* and stage of development, or other non-necessarily morphological factors. The choice of the OTU is obviously instrumental in determining what are the features that we should look at in order to assess similarities and differences between taxa. These *taxonomic characters* are *"a characteristic (or feature) of one kind of organism that will distinguish it from another kind"* (ibidem, p.71); in a morphological perspective, it is the character's variable phenotypic expression, or character *state*, that we assess either qualitatively or quantitatively in our attempt to establish similarities and differences between OTUs. As classification is based on comparison, when comparing two different OTUs in search of their level of similarity, what we really do is comparing the state of their characters. It is therefore a pre-requisite of any classification in Systematics that we do that *"over a set of characteristics applicable to both of them"* (ibidem, p.75), or, more precisely, over *homologous characters*. For example, we may want to establish what are the species represented in a collection of plants (objective of the classification). For that purpose we classify individual plants rather than, for example, groups or populations of plants; in this case, a choice regarding the scale of our observation (that of the organism) leads to the identification of the OTU (the individual plant). Remaining in the area of morphology, a preliminary observation may reveal that some plants have serrated leaf edges while others have regular ones. Being serrated or regular are discriminatory *states* of that particular character of the plant, the *leaf edge*, which is regarded to be *homologous* in the case in question.

In the transition from living organisms to cities, the Systematics approach encounters several problems, the most important of which pertains to the first step, the identification of the OTU. In biological systematics, classification at almost all levels is based on individual organisms, an entity that is in most cases unambiguous. That is not the case in urban studies where the criteria for the selection of the OTU must be elucidated in a far less intuitive manner. What is "the organism" in cities? Is it the city itself, or the district, the neighbourhood, or the street? Our urban morphometric analysis aims at: a) identifying the unit of analysis (OTU), b) rigorously defining the constituent elements of the urban form which, at the scale of the unit of analysis, are universally correspondent (homologous characters); c) determining the visible qualities that these elements can take in the real world (character state); d) adopting a system to quantitatively measure these visible qualities which is universally applicable and replicable; this must include the identification of the smallest set of variables able to deliver an *appropriate* description of cases, and a reliable validation theory against which such appropriateness is tested.

Finally, though the rigorous description and classification of organisms practiced in biological systematics must be regarded as fundamentally distinct from inferences of their ancestral relationships or common descent, which is specific to Phylogeny (Borgmeier, 1957, p. 54), our effort to establish an Urban Morphometrics discipline opens the way for further explorations of what we may evocatively call "the urban tree". In this paper, we derive a dendrogram that represents overall morphological similarity, not necessarily decent. However, according to MacLeod (MacLeod, 2002, p. 100), *"morphological data are regarded as being of significance in systematics because morphological variation is believed to be characterized by gaps between taxa. The presence of these gaps makes each taxon uniquely diagnosable and their hierarchical structure reflects action of morphological change superimposed on the evolutionary process of ancestry and descent. These gaps may arise as a result of a number of evolutionary processes, but their discovery, description, and interpretation represents the first and most basic task of all systematics research"*.

## 2. The Urban Morphometrics analysis of forty-five "sanctuary areas"

*2.1. Method*

*"That there is order in nature is a presupposition of any scientific research"* (Borgmeier, 1957, p. 53); however, *"nature is highly complex and the multiplicity of forms is oppressive"* (ibidem, p. 54). The co-presence in the real world of an inner structure that is permanent and universal, and of visible manifestations of endless diversity is the signature of life. *"Diversity and unity are the two underlying themes that seem to characterize all life"* (Savage, 1963, p. iii). Any classification is in essence the attempt to reproduce the more stable and recurrent part of the dualism that sits outside of us, in the real world. It is, therefore, *structural* in nature. Urban morphometrics is our attempt to understand (reproduce) the permanent and universal structure of cities, the one that lays the ground for the amazing diversity of their visible forms. This requires that we direct our attention not to what makes cities different from one another, but what makes them similar in the first place.

The first step in doing so it to determine the appropriate OTU. Our OTU must be: a) universally present in all cities; b) large enough to represent a complete spectrum of all constituent elements of urban form, such that their homologous characters can be rigorously defined and measured; c) small enough to be morphologically specific; d) functionally recognizable, at its own scale, in the organizational structure of the city. The *Sanctuary Area* (SA) is the portion of urban form enclosed by intersecting Urban Main Streets (Mehaffy et al., 2010) (Fig.2). The SA can be determined objectively, consistently and internationally (Porta, Romice, Maxwell, Russell, & Baird, 2014), therefore it complies to the criteria above and has been adopted as the OTU of this study, where we demonstrate that it is a significant constituent of the urban evolutionary processes. For the purpose of this paper, forty-five SAs are studied in 45 cities, 40 of which in the UK. The reader is advised that all the case studies are named after the city to which they belong, but they do only represent one SA within that city.

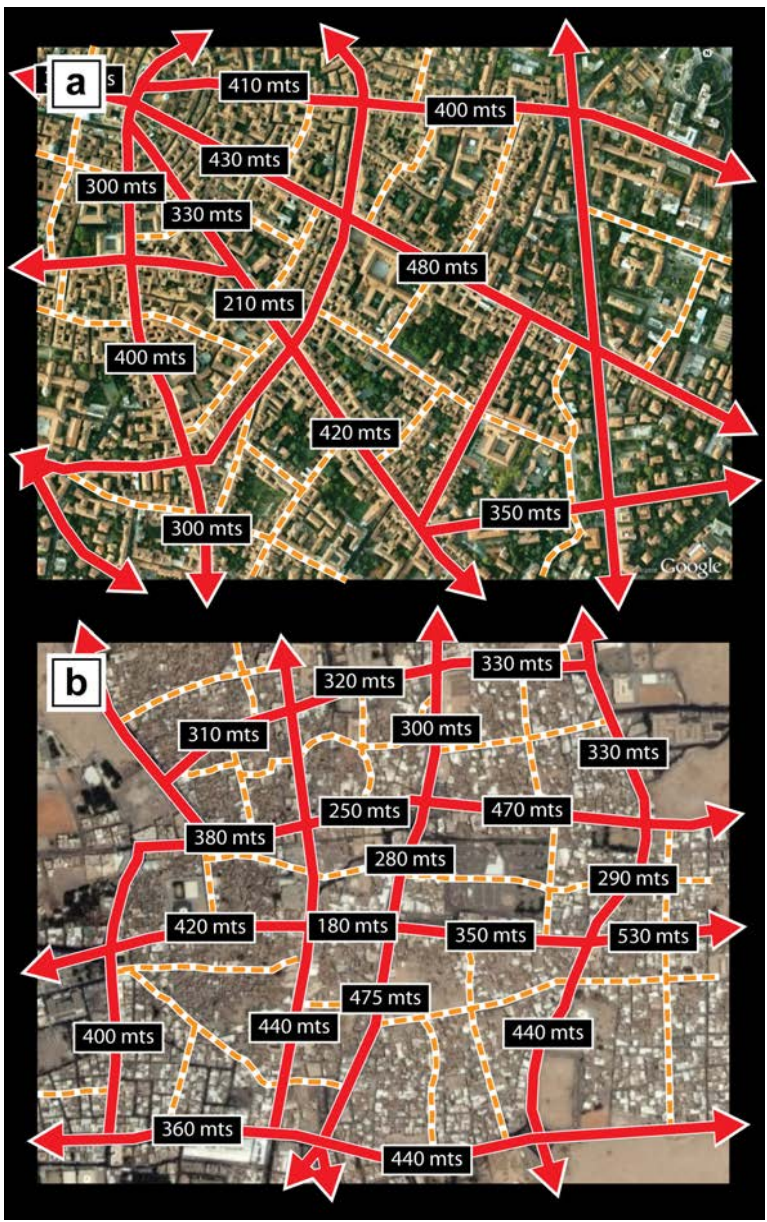

**Figure 2. The urban fabric defined by main streets: the Sanctuary Area**
The pattern of main streets (thick solid red) and sanctuary areas (on the background, defined by main streets) in the cities of Bologna, IT (a), and Al Hofuf, SA (b). Local main streets (dashed orange) emerge regularly within sanctuary areas as denser shortcuts connecting the higher network of main streets. Source: Mehaffy, Porta, Rofè, and Salingaros (2010).

The homologous characters of urban form must then be identified at the scale of the SA. These characters are features of the basic elements of urban form, in the same way that the leaf edge in a plant is a feature of one basic element of the plant, the leaf. Characters have been carefully selected according to these three main criteria: a) being significant features of the form and evolution of the urban fabric, as acknowledged in relevant literature; b) being detectable universally across all types of SAs of all cities, in all times; b) being measurable without direct inspection in situ, i.e. remotely through the most popular on-line mapping repositories such as Google Maps, OpenStreetMap and, in the UK, Ordnance Survey Maps. Many of these elements and characters are commonly recognised in urban morphology literature (Tab.1); however, some of them are innovative and are reflections of what best discriminates different types of urban developments over history. For example, we distinguished between *Regular Plots* and *Internal Plots*: whereby contrary to the latter, the former have a primary edge on, or access from, a public street. The rigor of a morphometrics approach has also required a significant work to redefine univocally, as much as possible, elements and characters of the taxonomic structure that have been so far extensively used in the literature, such as *street*, *block*, *plot*, *building density*, *built front*, *street edge* etc., whose definition has nevertheless so far been treated either informally or inconsistently, if not omitted altogether.

| OTU | Elements | | | Category of character |
|---|---|---|---|---|
| Sanctuary Area | | | | (1)(2)(3)(4)(5) |
| | Streets | | | (0) |
| | | Urban Mains | | (1)(5)(6) |
| | | Internal Streets | | (1)(5)(6) |
| | | | Local Mains | (1)(5)(6) |
| | | | Locals | (1)(5)(6) |
| | Blocks | | | (1)(2)(3)(4)(6) |
| | | Regular Plots | | (1)(2)(3)(5)(6) |
| | | Internal Plots | | (1)(2)(3)(5)(6) |
| | | Internal Ways | | (3) |
| | | Open Space | | (3) |
| | | Natural Areas | | (3) |

'Legend: (1) Size; (2) Shape; (3) Composition; (4) Density; (5) Usage; (6) Arrangement; (0) No metrics

| Category of character | Character | Unit of measure | Variable |
|---|---|---|---|
| (1) Size | Regular Plots Area | $m^2$ | Variables… |
| (2) Shape | Regular Plot Compactness Index | $m^2/m^2$ | Variables… |
| (2) Shape | Regular Plot Rectangularity Index | $m^2/m^2$ | Variables… |
| (3) Composition | Regular Plots per Block | n | Variables… |
| (5) Usage | Regular Plot Residential Use Ratio | $m^2/m^2$ | Variables… |
| (5) Usage | Regular Plot Non-Residential Use Ratio | $m^2/m^2$ | Variables… |
| (5) Usage | Regular Plot Mixed-Use Ratio | $m^2/m^2$ | Variables… |
| (5) Usage | Regular Plots Service Use Ratio | $m^2/m^2$ | Variables… |
| (5) Usage | Regular Plot Recreational Use Ratio | $m^2/m^2$ | Variables… |
| (6) Arrangement | Regular Plots Extension on Street | m | Variables… |
| (6) Arrangement | Regular Plot Covered Area Ratio | $m^2/m^2$ | Variables… |
| (6) Arrangement | Regular Plots per 100m Urban Mains | n/m | Variables… |
| (6) Arrangement | Regular Plots per 100m Local Mains | n/m | Variables… |
| (6) Arrangement | Regular Plots per 100m Local Streets | n/m | Variables… |

**Table 1. Taxonomic structure**
The complete structure is here presented only for the case of Regular Plots.

Ultimately, 75 homologous characters were extensively measured by means of 207 *variables*, spanning from, for example, the *built-front ratio* of the blocks to the *covered area ratio* of the plots, or the *ingress/egress ratio* of the SA. A total of 45 SAs were selected from an equal number of different cities, 40 of which are in the UK. Each SA was accurately mapped in a Geographic Information System (GIS) environment: all 2D characters such as Building coverage were identified spatially on the map, while 3D characters such as Building Height were added after inspection through Google Street View, or similar publicly available on-line repositories; the database relative to the SA was finally stored in a Microsoft Excel format. Once the overall database of all the 45 SAs was completed, an additional 5 further cases were prepared to be used as "unknown" cases: of these, 4 are European non-UK, and 1 (Tripoli) is an Arabic historical centre from north Africa.

This study has focused primarily on the establishment of a substantive method of measuring cities, with little regard to data mining, which has been approached in a rather conventional way by extensive manpower deployment. However, the method itself has been accurately designed to support further developments in areas such as remote sensing and big data as pertinent to urban morphology (Carneiro, Morello, Voegtle, & Golay, 2010). This applies across the board to all phases of the research production. For example, all information utilized in this research is achieved remotely, without direct site analysis, and all procedures of data management and treatment have been brought to a standard where automation could be directly applicable.

In traditional systematic biology the homologous characters and character states of an organism are often identified with reference to organs that are easy to capture: for example, there is no confusion between wings and beaks in birds and straightforward linear measurements of distance such as length, which are typical of traditional morphometrics, are equally unequivocal. However, as illustrated above, in urban morphology any assertion regarding the scale of the OTU, its characters and character states needs to be tested and validated against a clear set of criteria. It is important though that such validation theory is in some way readily available to the common sense as much as the distinction between a wing and a beak is. In our study, we propose the validation of our system be the *historical origins* of the case studies.

It is common knowledge, as much as a matter of intensive scholarship since the dawn of modern urban morphological studies (Poëte, 1924-1931), that the historical origin of an urban area has a direct and enduring impact on its evolution over time. The complex intricacies of technological, cultural, social and environmental factors that are conducive to a certain way of laying out streets, plots and buildings are all historically specific and converge into the production of the built environment in quite easily recognizable ways, so that, for example, we can intuitively distinguish, even after centuries, medieval from industrial parts of towns, and equally industrial from post-war suburban sprawl. What distinguishes urban fabrics of different historical origins in all evidence goes beyond factors of architectural language or style, and appears to be inherent to their long-lasting morphological *structure*. For example, there is evidence that the street layout is among the most resilient components of urban form, as well as the plot structure, which is directly linked with it (Moudon, 1986; Strano, Nicosia, Latora, Porta, & Barthélemy, 2012). These have a direct influence on other crucial elements such as street centrality, building types, density and land uses (Caniggia & Maffei, 2001, c.1979). As it is this morphological structure that we want to ultimately capture with our classification, we need to establish a system of measurements that allows urban form to be classified in taxa that are *distinct* in terms of the historical period when they were initially established.

For this study, we identified highly distinguishable historical origins groups (as described in literature), in order to make the test as divisive as possible and reduce the likelihood of errors. These are: a) *Historical* (compact medieval town centres); b) *Industrial* (compact dense working class housing from the late 19th and early 20th century); c) *New Towns* (post-war "towers-in-the-park" modernist estates); d) *Sprawl* (post-war low density and low rise "lollipop" suburbs). The four historical origin groups also belong to the two higher taxonomic levels of pre and post war developments, and are representative of clearly distinct building traditions and urban design models that are nevertheless common to much of the Western World at the very least, and especially to the UK as a whole. The decision of which SAs to be considered was informed by an extensive literature review. Cases were only included if they: a) were widely acknowledged in the literature to be representative of the typical form of their time of origin, and b) demonstrated in their contemporary appearance a reasonably homogeneous expression of that form across the entire case (Tab.2). All cases in fact, no matter their historical origins, are contemporary living urban environments in all respects ("living organisms"). These four historical building origins are quite distinct and are incorporated into this study to underpin our validation theory. If the Systematics approach adopted is sufficient to distinguish *between* these four groups, and yet identify similarities *within* the groups, then there is sufficient evidence that in fact the OTU, scale, characters and metrics utilized are appropriate. These claims are validated through several multivariate statistical analyses that are presented in the next section.

| N. | Origin Group | Sanctuary Area | Country | N. | Origin Group | Sanctuary Area | Country |
|---|---|---|---|---|---|---|---|
| 1 | Historic | Aberystwyth | Wales, UK | 24 | *New Town* | *Albertslund* | *Denmark* |
| 2 | Historic | Berwick-upon-Tweed | England, UK | 25 | New Town | Basildon | England, UK |
| 3 | Historic | Caernarfon | Wales, UK | 26 | New Town | Cumbernauld | Scotland, UK |
| 4 | Historic | Carlisle | England, UK | 27 | New Town | East Kilbride | Scotland, UK |
| 5 | *Historic* | *České Budějovice* | *Czech Republic* | 28 | New Town | Glenrothes | Scotland, UK |
| 6 | Historic | Chester | England, UK | 29 | New Town | Harlow | England, UK |
| 7 | Historic | Chichester | England, UK | 30 | New Town | Hatfield | England, UK |
| 8 | Historic | Conwy | Wales, UK | 31 | New Town | Livingston | Scotland, UK |
| 9 | Historic | Edinburgh | Scotland, UK | 32 | New Town | Milton Keynes | England, UK |
| 10 | Historic | Norwich | England, UK | 33 | New Town | Runcorn | England, UK |
| 11 | *Historic* | *Tripoli* | *Libya* | 34 | New Town | Skelmersdale | England, UK |
| 12 | Historic | York | England, UK | 35 | Sprawl | Balloch (Inverness) | Scotland, UK |
| 13 | *Industrial* | *Berlin* | *Germany* | 36 | Sprawl | Blythe Bridge | England, UK |
| 14 | Industrial | Bolton | England, UK | 37 | Sprawl | Boston Spa | England, UK |
| 15 | Industrial | Castleford | England, UK | 38 | Sprawl | Dudsbury | England, UK |
| 16 | Industrial | Glasgow | Scotland, UK | 39 | *Sprawl* | *Lyon* | *France* |
| 17 | Industrial | Leicester | England, UK | 40 | Sprawl | Milltimber | Scotland, UK |
| 18 | Industrial | Liverpool | England, UK | 41 | Sprawl | Newton Mearns | Scotland, UK |
| 19 | Industrial | Manchster | England, UK | 42 | Sprawl | Penyrheol | Wales, UK |
| 20 | Industrial | Middlesbrough | England, UK | 43 | Sprawl | Syston | England, UK |
| 21 | Industrial | Newcastle-upon-Tyne | England, UK | 44 | Sprawl | Upton | England, UK |
| 22 | Industrial | Preston | England, UK | 45 | Sprawl | Winterbourne | England, UK |
| 23 | Industrial | Skipton | England, UK | | | | |

(*) Non-UK cases in *italics*

**Table 2. List of cases and their historical origins.**
Cases are Sanctuary Areas nominated after the city they belong to.

*2.2. Findings*

*Principal Components Analysis* (*PCA*) is one of the oldest and most largely used techniques for multivariate data analysis (Hair, Black, Babin, Anderson, & Tatham, 2006). It is widely employed by statisticians in a range of disciplines and is applicable in many scientific studies with various types of data. As a form of *Exploratory Data Analysis* (*EDA*), it is used at a preliminary stage in statistical analyses to reveal whether there are any groupings in the data, outliers or dominant trends (Brereton, 2009). *PCA* aims at reducing a set of observations characterized by a large number of possibly correlated variables into a set of values characterized by a smaller number of uncorrelated Principal Components, yet accounting for a sufficient amount of the variability in the data. These Principal Components represent linear combinations of the original variables and can be considered as variables themselves in *EDA*. Therefore, rather than trying to understand the behaviour of the data measured against 45 case studies in 207 dimensions (the number of variables utilized to measure our 45 SAs), *PCA* allows for a much more straightforward analysis, in fewer dimensions, utilising only the first few Principal Components. The *PCA* makes it possible to reveal the underlying characteristics and relationships in the structure of the data, in a way that is straightforward to observe graphically in two or three-dimensional charts. In our case (Fig.3), *PCA* allows us to make two important observations: first, that the 45 cases comfortably cluster according to their historical origins, which satisfies the validation theory; and second, that a quite sharp distinction emerges between pre and post war cases with respect to the selected Principal Components.

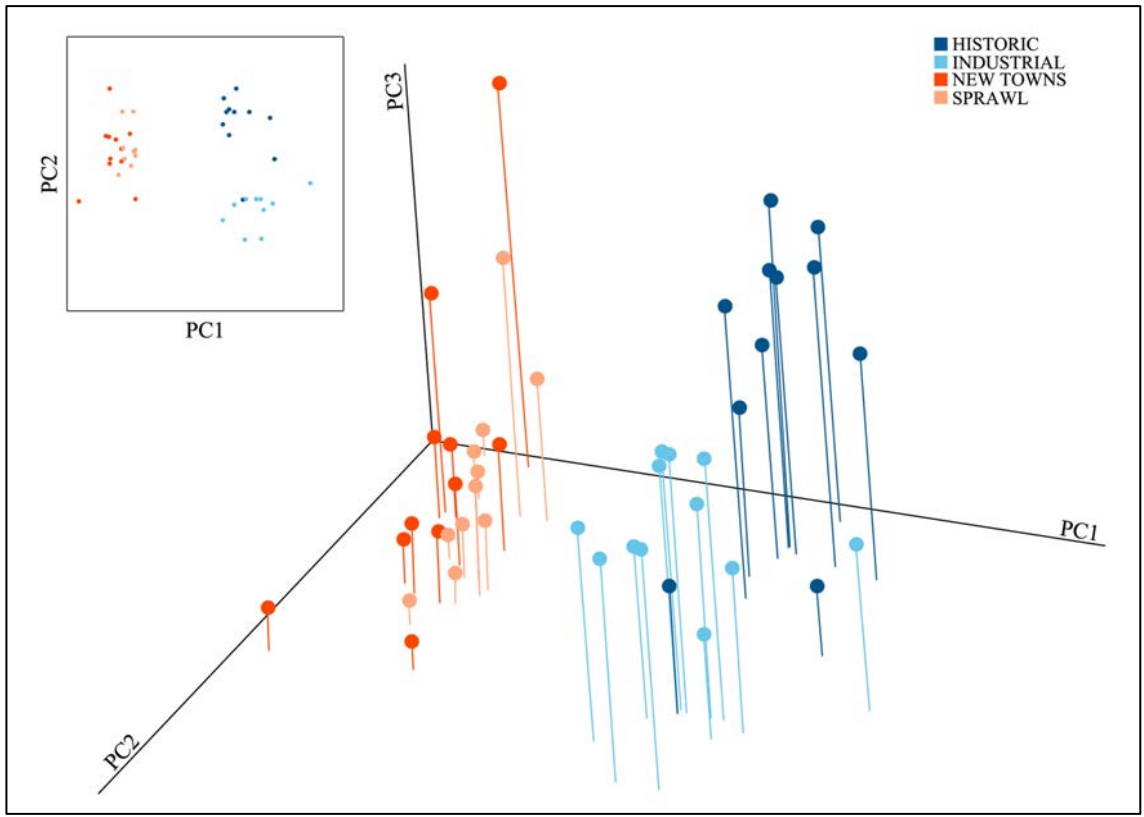

**Figure 3. Principal Component Analysis (*PCA*) of the 45 sanctuary areas**
The Principal Component Analysis reveals that there is sufficient information to describe the underlying behaviour of the data in as few as the first three Components alone. PC1 accounts for a neat separation between cities with pre or post war origins and PC2 acts in separating between the four origin groups. There is also information held in PC3 contributing in the separation between origin groups, especially between New Towns and Sprawl. The 3D scores plot shows these three tiers of separation and the 2D scores plot gives a very clear view of the inherent groupings of the case studies.

A Cost Benefit Analysis (*CBA*) is developed for this study in order to analyse the relative benefit of including more variables, in a parallel effort to reveal which variables are most important in the morphometric analysis of urban form; with "most important" we mean, in the context of this research, most *discriminatory* (Fig.4): in short, we explore what is the contribution of each variable in distinguishing cases according to the four historical origin groups. *CBA* utilises the *Fisher Weight* measure to rank the variables based on their overall discriminatory ability between the four origin groups. *CBA* proceeds iteratively to test for correct classification (using a *Linear Discriminant Analysis*) of cases when analysed using only the first top-ranked variable, then the first and the second, and so on (Brereton, 2009). This proceeds iteratively for the top 100 variables repeated for 100 test and training data set splits, representing one third and two thirds of the total case studies, respectively. We observe that the variables ranked in the top 9 positions of *CBA* (Tab.3) allow for over 90% average correct classification rate in relation to the four origin groups. These top variables make evidence of the extent to which buildings line up in close proximity to the block's perimeter as opposed to showing significant setback, and how this phenomenon occurs over local rather than main streets; or the way buildings are laid out within the block, either covering much of it or just a little; or the extent to which regular plots are characteristic of the SA's composition as opposed to internal plots. It is worth noting that the high discretionary capacity of such characters seems to express quite neatly the inherent opposition that has marked the intellectual history of urban design models as applied in particular to the ordinary urban environment, in the crucial passage from the pre-modern age of master-builders to that of advanced artists/professionals, or "Palladio's children" (Habraken, 2005). In particular, these characters distinguish the traditional compact urban form from the various post-war expressions of the garden city and the towers-in-the-park models (Hall, 2002, c.1988).

| CBA Ranking | Character | Variable |
|---|---|---|
| 1 | Block Built Front Ratio | Interquartile Average |
| 2 | Block Covered Area Ratio | Interquartile Average |
| 3 | Block Covered Area Ratio | Maximum |
| 4 | Local Street Built Front Ratio | Interquartile Average |
| 5 | Block Built Front Ratio | Maximum |
| 6 | Regular Plot Covered Area Ratio | Interquartile Average |
| 7 | Urban Main Built Front Ratio | Interquartile Average |
| 8 | Sanctuary Area Regular Plot Ratio | Ratio |
| 9 | Urban Main Built Front Ratio | Maximum |

**Table 3. List of the 9 top-ranked variables in the Cost Benefit Analysis.**
Their cumulative representation of the variability in the data overall goes beyond the 90%.

Interestingly, in respect to correct classification rates between pre and post war fabric, there is 100% correct classification regardless of the number of variables considered. Moreover, we show in the inset of fig.4 the scores plots of the first two Principal Components resulting from only the 9 top-ranked variables: the visible separation between the four groups is still quite strong, except for a few outliers that can in fact be easily explained by looking at their specific form.

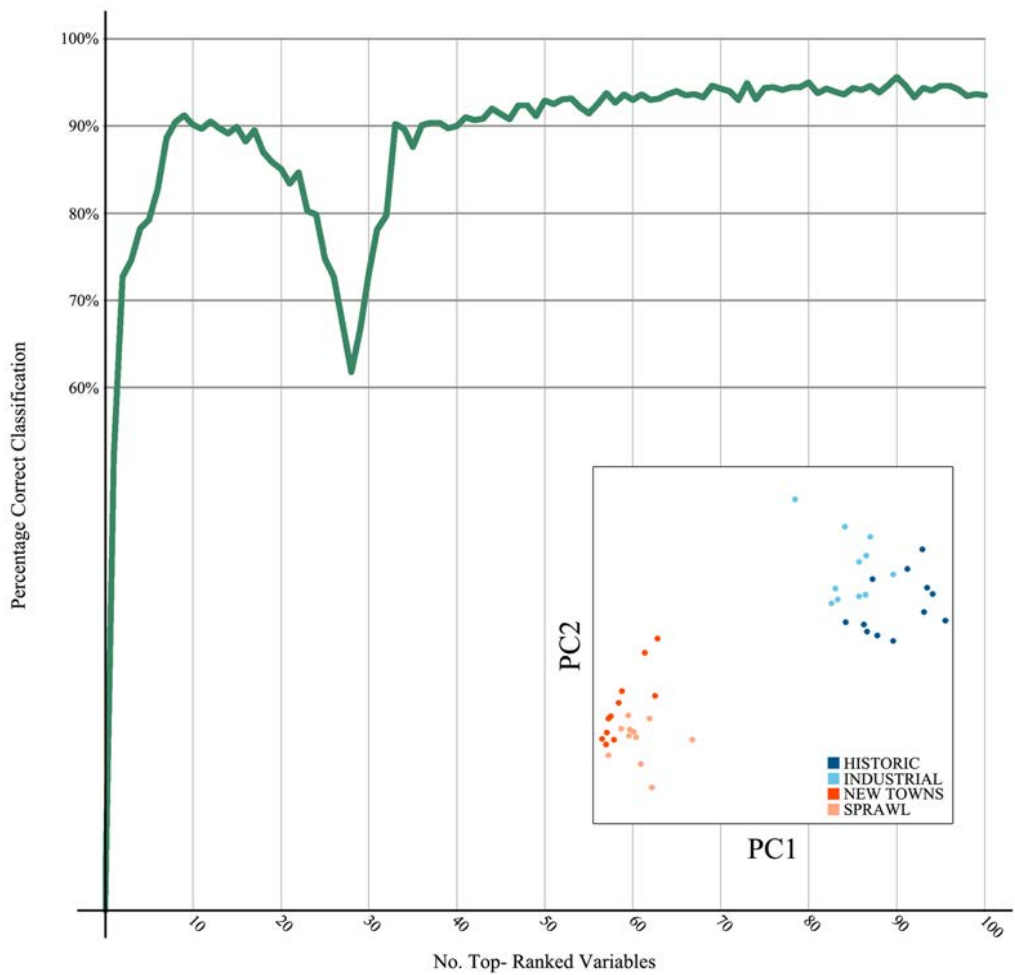

**Figure 4. Cost Benefit Analysis (*CBA*) of the 45 sanctuary areas**
The Cost/ Benefit Analysis applied in this study incorporates the Fisher Weight test of overall discriminatory ability of the 207 variables in distinguishing between origin groups. Essentially, this reveals the most important metrics of form and orders them. With only 9 variables, an average percentage of correct classification of over 90% was demonstrated. The two-dimensional *PCA* scores plot in the inset shows that with only 9 variables there is still a remarkably clear distinction between pre and post-war developments. Compared with the *PCA* of 207 variables shown in Fig.3 (inset), there is not as strong of separation between the two pre-war origin groups, which in any case still form compact and distinct clusters. However, the distinction between the post-war origin groups has improved as the clusters have become slightly more compact and distinct.

*Hierarchical Clustering Analyses* (*HCA*) are common methods of visually expressing the relationship between OTU's and are common in Systematics studies (Gordon, 1996; Legendre & Legendre, 1998). We show a dendrogram representing the relationship between the 45 original cases (Fig.5) measured with only the 9 variables that were ranked highest in the *CBA*; the distance along the X axis of the points at which the cases are joined represents their grade of similarity, where the closer the branches join to the left the more similar they are. Utilizing *Ward's* method and considering the *Euclidean* distance between cases, the dendrogram has a *Silhouette Coefficient*, a test of goodness of fit of the clustering to the data, of 0.49, demonstrating the reliability of the dendrogram (Kaufman & Rousseeuw, 2005). The grouping of the cities into the anticipated groups upholds the validation theory proposed at the beginning of this study. Moreover, the remarkably large split between pre and post war groups emerges, supporting the idea that something happened to our cities after WWII that has for the first time in urban history affected the fundamental structure of their form.

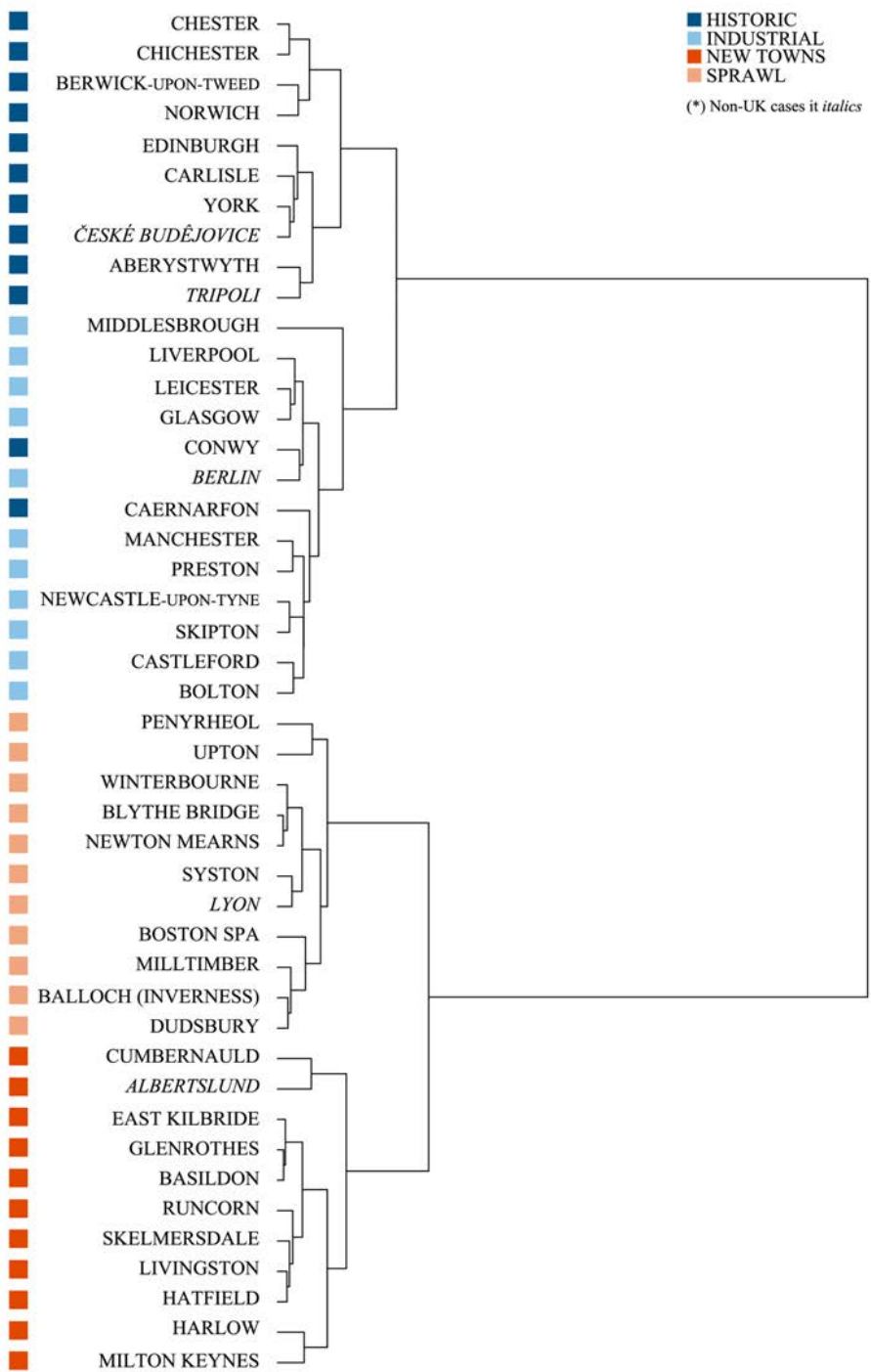

**Figure 5. Dendrogram of the 45 sanctuary areas**
Considering the reduced data set of only the 9 top-ranked variables, a Hierarchical Clustering Analysis using Ward's Method and Euclidean distance to estimate resemblance produces this dendrogram. The closer two cities are joined to the left, the more overall similarity they share. Barring a few of the known outliers, the dendrogram demonstrates near perfect classification of the four origin groups an absolutely perfect separation between pre and post-WW2 cities. Further, the international case studies are nicely classified, furthering supporting the robustness of the morphometrics method, even in its reduced variable set form.

# 3. Conclusions

In this paper we present a research on 45 "sanctuary areas" of predominantly UK cities. A sanctuary area is defined as a portion of the urban fabric that is bounded by main streets. The research focuses on the urban form of such portions, and is aimed at establishing a methodology of analysis of its characteristics and their relationship that is rigorous, quantitative, universally applicable and could ultimately be useful in city-making. By doing so, we claim that we are opening a new avenue of research in urban morphology that we name *urban morphometrics*, following the analogy with biological morphometrics. Our research suggests that urban morphology as a field of knowledge exhibits a clear gap in this area, and that this gap has prevented a richer cross-fertilization with evolutionary biology, beyond the simple analogy.

However, we acknowledge that such an ambitious programme of research requires long-term plans, and that we are with this paper only making the first step towards it. As such, taking lessons from biological systematics, our aim in this research is primarily that of establishing a method for the classification of urban form. For this purpose, we propose here the fundamental elements of a systematic method of comparative urban morphometrics, and we apply it to the aforementioned 45 real-world cases of contemporary urban fabrics. We then undertake a rigorous statistical analysis of the dataset aimed at: a) verifying that the method captures the historical origins of cases by correctly classifying them according to four predetermined historical groups (*historical*, *industrial*, *new towns* and *sprawl*); b) identifying the minimal set of variables that "explain" sufficiently the variability of the data, i.e. those variables that are indispensible to correctly classify cases according to their historical origins; c) producing a first "urban tree" that expresses the hierarchical grouping of cases according to their morphological similarity.

Findings show that: a) overall, the method appears to perform very well in clustering cases in a way that highly correlates with their historical origins; b) it is possible to derive a very neat hierarchical representation of the cases' similarity by using just the 9 variables previously mentioned; c) the great divide between pre and post-WWII cases suggests that we may be witnessing there a phenomenon of remarkable evolutionary significance. Though this representation expresses only the morphological *similarity* between cases, and does not *per se* introduce conclusions of phylogenetic nature, we suggest that this first result is encouraging both in terms of robustness of the method and fertility towards further progresses in different areas of science, including phylogeny.

Moreover, we find four major directions for the further development of this work. Firstly, urban morphometrics must be tested at a much larger scale; that implies the development of a reliable protocol of data mining that takes advantage of technologies of remote sensing and image processing over big data on-line repositories. Secondly, the definition of the sanctuary area as the Operational Taxonomic Unit (OTU) requires a much deeper investigation of their profile, for example in terms of their organizational, developmental, regulatory, functional and emergent non-reducible features (Savage, 1963, p. 12). Thirdly, reflection must be put in the multi-scalar nature of cities, of which the scale of the sanctuary area represents one level. Finally, significant work must be undertaken before a reliable and universally accepted set of characters and variables can be considered achieved even just at the scale of the sanctuary area; further investigation is needed in particular to distinguish finer-grained taxa within the two camps of pre and post war urban fabrics, and even within the level of the four origin groups identified in this study.

# References

Adams, D. C., Rohlf, F. J., & Slice, D. E. (2013). A field comes of age: geometric morphometrics in the 21st century. *Hystrix, the Italian Journal of Mammalogy, 24*(1), 7-14.

Alexander, C. (2004). *Sustainability and morphogenesis: The birth of a living world*. Paper presented at the Schumacher Lecture, Bristol, UK.

Bookstein, F. L. (1993). A brief history of the morphometric synthesis. In L. F. Marcus, E. Bello, & A. García Valdecasas (Eds.), *Contributions to morphometrics* (Vol. 8, pp. 15-40). Madrid: GRAFICAS MAR-CAR, S. A.

Borgmeier, T. (1957). Basic questions of systematics. *Systematic Zoology, 6*(2), 53-69.

Brereton, R. (2009). *Chemometrics for pattern recognition*: John Wiley & Sons.

Caniggia, G., & Maffei, G. L. (2001, c.1979). *Architectural Composition and Building Typology: Interpreting Basic Building*. Florence: ALINEA.

Carneiro, C., Morello, E., Voegtle, T., & Golay, F. (2010). Digital urban morphometrics: automatic extraction and assessment of morphological properties of buildings. *Transactions in GIS, 14*(4), 497-531.

Cataldi, G., Maffei, G. L., & Vaccaro, P. (2002). Saverio Muratori and the Italian school of planning typology. *Urban Morphology, 6*(1), 3-14.

Conzen, M. P. (2001). The study of urban form in the United States. *Urban Morphology, 5*(1), 3-14.

Conzen, M. R. G. (1960). Alnwick, Northumberland: a study in town-plan analysis. *Transactions and Papers (Institute of British Geographers)*, iii-122.

D'Arcy Thompson, W. (1942, c.1917). *On growth and form*. New York: McMillan.

Darin, M. (1998). The study of urban form in France. *Urban Morphology, 2*(2), 63-76.

Dawkins, R. (2006, c.1976). *The selfish gene*: Oxford university press.

Gordon, A. D. (1996). Hierarchical classification. In P. Arabie, L. J. Hubert, & G. De Soete (Eds.), *Clustering and classification* (pp. 65-121): World Scientific.

Gu, K., & Zhang, J. (2014). Cartographical sources for urban morphological research in China.

Habraken, N. (2005). Palladio's Children, seven essays on everyday environment and the architect. Edited by Jonathan Teicher, Editor: Oxford UK, Taylor & Francis.

Hair, J. F., Black, W. C., Babin, B. J., Anderson, R. E., & Tatham, R. L. (2006). *Multivariate data analysis* (Vol. 6): Pearson Prentice Hall Upper Saddle River, NJ.

Hall, P. (2002, c.1988). *Cities of tomorrow: an intellectual history of urban planning and design in the twentieth century*: Blackwell Publishing.

Heywood, V. H. (1976). *Plant taxonomy*: London: Edward Arnold 63p.(Institute of Biology's Studies in Biology no. 5)-Illus.. General (KR, 197600023).

Ibarz, J. V. (1998). The study of urban form in Spain. *Urban Morphology, 2*(1), 35-44.

Jeffrey, C., & Heywood, V. H. (1977). *Biological nomenclature*: Edward Arnold London.

Kaufman, L., & Rousseeuw, P. J. (2005). Finding groups in data: an introduction to cluster analysis. 2005: John Wiley & Sons, Inc.

Klingenberg, C. P. (2002). Morphometrics and the role of the phenotype in studies of the evolution of developmental mechanisms. *Gene, 287*(1), 3-10.

Legendre, P., & Legendre, L. (1998). Numerical ecology: second English edition. *Developments in environmental modelling, 20*.

MacLeod, N. (2002). Phylogenetic signals in morphometric data. *Morphology, shape and phylogeny*, 100-138.

Manktelow, M. (2010). History of taxonomy. *Lecture from Dept. of Systematic Biology, Uppsala University. DOI=* http://atbi. *eu/summerschool/files/summerschool/Manktelow_Syllabus. pdf*.

Marshall, S. (2008). *Cities Design and Evolution*: Routledge.

Mehaffy, M., Porta, S., Rofè, Y., & Salingaros, N. (2010). Urban nuclei and the geometry of streets: The 'emergent neighborhoods' model. *Urban Design International, 15*(1), 22-46. doi: 10.1057/udi.2009.26

Moudon, A. V. (1986). *Built for change: neighborhood architecture in San Francisco*: Mit Press.

Muratori, S. (1960). *Studi per una operante storia urbana di Venezia*. Roma: Istituto Poligrafico dello Stato.

Pagel, M. (2012a). Evolution: adapted to culture. *Nature, 482*(7385), 297-299.

Pagel, M. (2012c). *Wired for culture: The natural history of human cooperation*: Penguin UK.

Poëte, M. (1924-1931). *Une vie de cité: Paris de sa naissance à nos jours*: A. Picard.

Porta, S., Romice, O., Maxwell, J. A., Russell, P., & Baird, D. (2014). Alterations in scale: Patterns of change in main street networks across time and space. *Urban Studies, 51*(16), 3383-3400. doi: 10.1177/0042098013519833

Reyment, R. A. (2010). Morphometrics: an historical essay *Morphometrics for Nonmorphometricians* (pp. 9-24): Springer.

Richerson, P. J., & Boyd, R. (2008). *Not by genes alone: How culture transformed human evolution*: University of Chicago Press.

Rohlf, F. J., & Marcus, L. F. (1993). A revolution morphometrics. *Trends in Ecology & Evolution, 8*(4), 129-132.

Roth, V. L., & Mercer, J. M. (2000). Morphometrics in development and evolution. *American Zoologist, 40*(5), 801-810.

Savage, J. M. (1963). *Evolution*. New York: Holt, Rinehart and Winston.

Siksna, A. (2006). The study of urban form in Australia. *Urban Morphology, 10*(2), 89.

Simpson, G. G. (1961). *Principles of animal taxonomy*: Columbia University Press.

Sneath, P. H., & Sokal, R. R. (1973). *Numerical taxonomy. The principles and practice of numerical classification*: W.H.Freeman & Co Ltd.

Steadman, P. (2008). *The Evolution of Designs: Biological analogy in architecture and the applied arts*: Routledge.

Strano, E., Nicosia, V., Latora, V., Porta, S., & Barthélemy, M. (2012). Elementary processes governing the evolution of road networks. *Scientific Reports, 2*. doi: doi:10.1038/srep00296

Whitehand, J. W. (2001). British urban morphology: the Conzenion tradition. *Urban Morphology, 5*(2), 103-109.